# Maxwell-compensated design of asymmetric gradient waveforms for tensor-valued diffusion encoding

Filip Szczepankiewicz[1,2], Carl-Fredrik Westin[1,2] and Markus Nilsson[3]

1. Radiology, Brigham and Women's Hospital, Boston, MA, US
2. Harvard Medical School, Boston, MA, US
3. Clinical Sciences, Lund, Lund University, Lund, Sweden

*Corresponding author*

Filip Szczepankiewicz

E-mail: fszczepankiewicz@bwh.harvard.edu

Address: 1249 Boylston St. Boston, MA, USA

Telephone: +46763194499

**Abstract**

Purpose: Asymmetric gradient waveforms are attractive for diffusion encoding due to their superior efficiency, however, the asymmetry may cause a residual gradient moment at the end of the encoding. Depending on the experiment setup, this residual moment may cause significant signal bias and image artifacts. The purpose of this study was to develop an asymmetric gradient waveform design for tensor-valued diffusion encoding that is not affected by concomitant gradient.

Methods: The "Maxwell index" was proposed as a scalar invariant that captures the effect of concomitant gradients and was constrained in the numerical optimization to 100 $(mT/m)^2$ms to yield "Maxwell-compensated" waveforms. The efficacy of this design was tested in an oil phantom, and in a healthy human brain. For reference, waveforms from literature were included in the analysis. Simulations were performed to investigate if the design was valid for a wide range of experiments and if it could predict the signal bias.

Results: Maxwell-compensated waveforms showed no signal bias in oil or in the brain. By contrast, several waveforms from literature showed gross signal bias. In the brain, the bias was large enough to markedly affect both signal and parameter maps, and the bias could be accurately predicted by theory.

Conclusion: Constraining the Maxwell index in the optimization of asymmetric gradient waveforms yields efficient tensor-valued encoding with concomitant gradients that have a negligible effect on the signal. This waveform design is especially relevant in combination with strong gradients, long encoding times, thick slices, simultaneous multi-slice acquisition and large/oblique FOVs.

# Introduction

Diffusion-weighted MR imaging employs magnetic field gradients to sensitize the signal to the diffusive motion of signal carrying spin. The gradients are assumed to be linear in space, but in practice, they are always accompanied by so-called *concomitant gradients*, also referred to as *Maxwell terms*[1]. In spin-echo sequences, diffusion-weighting is most commonly achieved with an identical pair of pulsed field gradients on either side of the refocusing pulse. This configuration creates an elegant symmetry that ensures that concomitant gradient effects created by the first pulse are cancelled by the second. This means that, although the desired gradient waveform will be slightly perturbed by concomitant gradients, the $0^{th}$ moment of the actual gradient waveform will be zero, as is desired.

The same compensation of concomitant gradients cannot be assumed to be true for asymmetric gradient waveforms. Instead, asymmetric waveforms are likely accompanied by a spatially dependent concomitant gradients that cause non-negligible residual moments at the end of the encoding. Depending on the setup of the imaging sequence, a residual moment causes a shift in k-space that may manifest as through-slice dephasing, image blurring[2], $T_2^*$-relaxation[3], and signal-dropout[4], all of which reduce the accuracy of quantities or imaging biomarkers estimated from the data. Despite this, asymmetric waveforms are an attractive alternative to symmetric designs because of their superior efficiency. For example, asymmetry can be exploited in experiments that yield encoding periods of unequal duration (asymmetric timing) where all available time can be used for diffusion encoding, rather than inserting periods where the gradients are turned off[5,6]. This facilitates a categorical improvement of the encoding efficiency (encoding strength per unit time) and can be used to reduce echo-times and the signal to noise ratio (SNR). Moreover, asymmetric designs have been used to reduce eddy-currents[7], enable motion and acceleration compensation[8] and facilitate pore shape estimation[9,10].

Concomitant gradients are a well-known pitfall for applications that rely on strong gradients, and their effects have been described in several contexts[1,2,4,11-15]. To our knowledge, two strategies have been applied to address the effects of concomitant fields in the context of diffusion-weighted imaging. The first is to correct the error by calculating the concomitant gradients at some location in space and subtract them from the desired waveform[2,4]. This approach corrects the errors in a single position in space, but data acquired at some distance from the selected point may still suffer from the error. The second strategy is to design waveforms that can be asymmetric but where the gradient moment from the concomitant gradients match on both sides of the refocusing pulse, or so-called *waveform reshaping*[16]. This approach has been applied for linear diffusion encoding (along a single spatial direction), however, it does not generalize to single-shot diffusion encoding along multiple directions.

We refer to single-shot diffusion encoding along multiple spatial directions as "tensor-valued" since it cannot be described by just a b-value and a direction vector, but is instead described by a b-tensor that also captures the *shape* of the encoding[17,18]. Tensor-valued encoding can probe features of the tissue microstructure that cannot be probed by conventional encoding alone[19-27], e.g. microscopic anisotropy, orientation coherence and isotropic heterogeneity of tissue[28]. It can also inform biophysical models[29].

Waveforms that yield tensor-valued encoding have been proposed in both symmetric and asymmetric variants[30-33]. Recently, Sjölund et al.[5] proposed a numerical optimization technique that can generate arbitrary b-tensor shapes with asymmetric waveforms that enabled a significant reduction of encoding times and improved SNR compared to previous designs.

In this work we propose a waveform design that negates effects of concomitant fields while enabling arbitrary shapes of the b-tensor. We propose the scalar *Maxwell index* as a cost function for minimization during waveform optimization and show that Maxwell-compensated waveforms do not suffer errors due to concomitant gradients.

**Theory**

*Approximation of concomitant gradients*

The Maxwell equations dictate that linear magnetic field gradients, such as those used for diffusion encoding, are accompanied by spatially dependent concomitant gradients[11]. The concomitant gradients depend on the main magnetic field ($B_0$) and the *desired* (subscript D) gradient waveform

$$\mathbf{g}_D(t) = \mathbf{R}\,[g_1(t)\ \ g_2(t)\ \ g_3(t)]^T = [g_x(t)\ \ g_y(t)\ \ g_z(t)]^T, \qquad \text{Eq. 1}$$

where $\mathbf{R}$ is a rotation matrix that determines how the three-dimensional waveform shape ($g_{1,2,3}$) is translated to the physical gradient axes ($g_{x,y,z}$). To abbreviate the notation, we will only use the latter format, but we note that all subsequent expressions depend on $\mathbf{R}$ even if not explicitly stated. The *concomitant* (subscript C) gradient waveform ($\mathbf{g}_C$) can be approximated with an expansion in $B_0$ at a given position ($\mathbf{r} = [x\ y\ z]^T$, z-axis is parallel with $B_0$)[1,11], according to

$$\mathbf{g}_C(t, \mathbf{r}) = \mathbf{G}_C(t)\mathbf{r}, \qquad \text{Eq. 2}$$

where the *concomitant gradient matrix* is

$$\mathbf{G}_C(t) \approx \frac{1}{4B_0}\begin{bmatrix} g_z^2(t) & 0 & -2g_x(t)g_z(t) \\ 0 & g_z^2(t) & -2g_y(t)g_z(t) \\ -2g_x(t)g_z(t) & -2g_y(t)g_z(t) & 4g_x^2(t) + 4g_y^2(t) \end{bmatrix}. \qquad \text{Eq. 3}$$

The *actual* (no subscript) gradient waveform, experienced by the spin, will not be the desired waveform, but rather the sum of the desired waveform and the concomitant waveform

$$\mathbf{g}(t, \mathbf{r}) = \mathbf{g}_D(t) + \mathbf{g}_C(t, \mathbf{r}). \qquad \text{Eq. 4}$$

Errors occurring due to concomitant fields appear because the actual gradient waveform is not balanced, i.e., is has a non-negligible 0[th] moment, which causes a shift in k-space. To capture the size of the k-space shift independent of the location we construct a *concomitant moment matrix*

$$\mathbf{K} = \frac{\gamma}{2\pi} \int_0^\tau h(t) \cdot \mathbf{G_C}(t) dt, \qquad \text{Eq. 5}$$

where $\gamma$ is the gyromagnetic ratio, and $\tau$ is the duration of the diffusion encoding. The sign function $h(t) = (-1)^{j(t)}$ represents the spin-dephasing direction and assumes values of 1 or –1 for each encoding period $j(t) \in \mathbb{N}$, such that neighboring periods—separated by a refocusing pulse—have opposite signs. This convention provides a straightforward generalization of this work to experiments with an arbitrary number of refocusing pulses.

The residual $0^{th}$ moment vector ($\mathbf{k}$) is the integral of the actual gradient waveform at any given position, according to

$$\mathbf{k(r)} = \frac{\gamma}{2\pi} \int_0^\tau h(t) \cdot \mathbf{g}(t, \mathbf{r}) dt \underbrace{= \mathbf{Kr}}_{\text{If balanced } \mathbf{g_D}}, \qquad \text{Eq. 6}$$

which can be simplified to the expression on the right-hand side if the desired waveform is balanced, i.e., if the desired $0^{th}$ moment $\mathbf{k_D} = \int_0^\tau h(t) \cdot \mathbf{g_D}(t) \, dt = 0$. Figure 1 visualizes the desired and concomitant gradient waveforms at multiple positions along with the resulting residual moment vectors. At the end of the gradient waveform, the residual $0^{th}$ moment ($k$) along an arbitrary direction ($\mathbf{n}$, unit vector) is a scalar that carries information about the local shift in k-space at position $\mathbf{r}$, according to

$$k(\mathbf{n}, \mathbf{r}) = \mathbf{n}^T \mathbf{Kr}. \qquad \text{Eq. 7}$$

In summary, concomitant gradients are proportional to the square of the gradient amplitude and inversely proportional to the main magnetic field strength (Eq. 3). The effects of concomitant gradients have a non-trivial dependency on rotation and spatial position.

*Signal error and artifacts caused by concomitant gradients*

Several spatially varying artifacts may manifest when $\mathbf{k} \neq 0$. For simplicity, we will describe the relevant effects for an ideal spin-echo sequence with rectangular slice profiles and echo-planar imaging (EPI) readout; these are slice dephasing and blurring[2] as well as $T_2^*$-weighting[3].

Slice dephasing is caused by a residual moment along the slice encoding direction ($\mathbf{n_S}$, unit vector normal to the slice plane), where the shift is $k_S(\mathbf{n_S}, \mathbf{r}) = \mathbf{n_S^T Kr}$ (Eq. 7). Assuming approximately linear phase dispersion within each voxel[12] and a rectangular slice profile[2], the slice attenuation factor ($AF_S$) is given by

$$AF_S(\mathbf{n_S}, \mathbf{r}) = \left| \text{sinc}(\mathbf{n_S^T Kr} \cdot T) \right|, \qquad \text{Eq. 8}$$

where $T$ is the slice thickness. Shifting the echo within the readout plane incurs blurring, as information about high spatial frequencies is shifted outside of the readout window. For EPI, any shift in the readout plane also incurs a spatially dependent $T_2^*$-weighting, whenever the intensity-generating readout is performed at different time than at the spin-echo echo time. This effect is most prominent for shifts along the phase direction due to the slower readout speed in that direction. The residual moment along the phase encoding direction ($\mathbf{n}_P$) is given by $k_P(\mathbf{n}_P, \mathbf{r}) = \mathbf{n}_P^T \mathbf{K} \mathbf{r}$. Assuming negligible $T_2^*$-relaxation during the relatively rapid readout of each k-space line, the attenuation factor ($AF_P$) can be approximated by the echo-shift in relation to the readout trajectory

$$\mathrm{AF}_P(\mathbf{n}_P, \mathbf{r}) = \exp\left(-\frac{|\mathbf{n}_P^T \mathbf{K} \mathbf{r}|}{\Delta k} \cdot \frac{\Delta t}{T_2^*}\right), \qquad \text{Eq. 9}$$

where $\Delta k = N/\mathrm{FOV}_P$ is the inverse distance between two acquired k-space lines, where $N$ is the parallel imaging factor and $\mathrm{FOV}_P$ is the field of view in the phase direction, and $\Delta t$ is the time between the acquisitions of two consecutive k-space lines. The total attenuation factor from both dephasing and $T_2^*$-weighting is the product

$$\mathrm{AF}(\mathbf{n}_S, \mathbf{n}_P, \mathbf{r}) = \mathrm{AF}_S(\mathbf{n}_S, \mathbf{r}) \cdot \mathrm{AF}_P(\mathbf{n}_P, \mathbf{r}). \qquad \text{Eq. 10}$$

We conclude that spatially dependent signal errors will be caused by concomitant fields and will manifest as deleterious signal attenuation (AF < 1). In addition to the waveform trajectory and the $T_2^*$ of tissue, these errors depend on several imaging parameters, such as slice position and angulation, in-plane acceleration, bandwidth and slice thickness.

*Signal bias independent of diffusion weighting*

In simulations we can calculate the relative signal bias as $\mathrm{RB} = \mathrm{AF} - 1$ because AF (Eq. 10) is independent of the diffusion-weighting. In practical measurements, we can separate the impact of concomitant gradients from the diffusion-weighting by comparing a candidate waveform to a reference waveform that is not affected by concomitant gradients; all other variables kept equal. In a substrate that exhibits purely Gaussian diffusion, it is enough to compare the signal across identical b-values to remove the effect of diffusion-weighting and retain only effects caused by the concomitant gradients. However, in the presence of microscopic or macroscopic diffusion anisotropy, the encoding must be matched with respect to the b-tensor. The b-tensor is calculated from the gradient waveform, according to

$$\mathbf{B} = \int_0^\tau \mathbf{q}(t)\mathbf{q}(t)^T dt, \qquad \text{Eq. 11}$$

where the dephasing q-vector is

$$\mathbf{q}(t) = \gamma \int_0^\tau \mathbf{g}_D(t') dt', \qquad \text{Eq. 12}$$

and the b-value is simply $b = \mathrm{Tr}(\mathbf{B})$. Notably, this definition ignores the concomitant gradients because they have a negligible effect on the b-tensor. With matching b-tensors for a given waveform (*i*) and its reference (ref), the relative signal bias can be estimated by comparing the two, according to

$$\text{RB} = \frac{S_i(\mathbf{B})}{S_i(0)} \cdot \frac{S_{\text{ref}}(0)}{S_{\text{ref}}(\mathbf{B})} - 1, \qquad \text{Eq. 13}$$

where negative values of the relative bias reflect that signal is lost (AF < 1). We note that this estimation assumes that other effects that may also depend on the exact shape of the waveform, such as, diffusion time, exchange, flow, eddy-currents are negligible.

**Methods**

*Maxwell index*

We seek a gradient waveform $\mathbf{g}_D$ for which $\mathbf{K} = 0$ independent of the waveform rotation so that the signal bias is negligible. We propose to achieve this condition by minimizing the "Maxwell index" ($m$), herein defined as

$$m = (\text{Tr}(\mathbf{MM}))^{\frac{1}{2}}, \qquad \text{Eq. 14}$$

where

$$\mathbf{M} = \int_0^\tau h(t)\mathbf{g}_D(t)\mathbf{g}_D^T(t)\mathrm{d}t. \qquad \text{Eq. 15}$$

We motivate this construct by noting that both $\mathbf{M}$ and $\mathbf{K}$ comprise the same self-squared and cross-terms found in $\mathbf{g}_D(t)\mathbf{g}_D^T(t)$, such that nulling the elements of $\mathbf{M}$ and/or $\mathbf{K}$ would satisfy our aim. In other words, $\mathbf{K} = 0$ if $\int_0^\tau h(t)g_i(t)\,g_j(t)\mathrm{d}t = 0$ for all axis combinations and waveform rotations. In the special case where the waveform is a one-dimensional trajectory ($g_{1D}$), this problem has a simple solution because $\mathbf{K} = 0$ when $\int_0^\tau h(t)g_{1D}^2(t)\,\mathrm{d}t = 0$ independent of rotation[16]. However, in the general case, $\mathbf{K}$ is not invariant to rotation of the gradient waveform and is therefore a poor target for optimization. Instead, we propose to base the optimization target on $m$ which is strictly positive and invariant to rotation. This means that for a given experimental setup, a waveform with a sufficiently small $m$ will yield negligible effects due to concomitant gradients independent of rotation, FOV-orientation, and position within the FOV. Note that the diagonal elements of $\mathbf{MM}$ are strictly positive so that $\text{Tr}(\mathbf{MM})$ is zero only if the diagonal elements are all zero, whereas $\text{Tr}(\mathbf{M})$ may be zero because $\mathbf{M}$ contains both positive and negative diagonal elements.

*Optimization of Maxwell-compensated waveforms*

Efficient tensor-valued diffusion encoding can be tailored to an arbitrary configuration of the imaging sequence and hardware capabilities using the numerical optimization framework by Sjölund et al.[5]. However, this framework was not designed to consider the effects of concomitant gradients.

We generate *Maxwell-compensated* waveforms (MCW) with arbitrary b-tensor shapes by including the Maxwell index (Eq. 14) as a constraint in the optimization framework[5]. To allow arbitrary rotation of the waveform, we used the Euclidean norm and a heat dissipation factor of 1.

In principle, the waveforms can be optimized at the temporal resolution employed by the gradient amplifier system. However, with the current implementation, the optimization time increases rapidly with the number of samples along the time dimension ($N$). Therefore, optimization was limited to a relatively coarse temporal resolution and linear interpolation was used to resample the waveform to an appropriate duration at the gradient raster time used by the gradient amplifier. Although this approach facilitates fast optimization of waveforms, linear interpolation does not perfectly preserve Maxwell-compensation. We circumvent this problem by predicting the signal error after interpolation in a worst-case scenario and increasing the temporal resolution until the error is negligible. For the purposes of this paper, the worst-case scenario is taken to mean that we use the actual imaging parameters of the sequence but considers the rotation of the waveform that gives the largest signal error within a sphere with a diameter of 0.5 m (centered on the isocenter) and a $T_2^*$ of 40 ms. Under these conditions, we consider a signal bias magnitude below 1 % (AF > 0.99) at the highest b-value to be negligible. Figure 2 shows that $N = 100$ is a sufficient temporal resolution for the waveform optimization and that the optimization should constrain the Maxwell index to be less than approximately 1000 (mT/m)$^2$ms. To create additional headroom, we constrain all Maxwell-compensated waveforms to have $m \leq 100$ (mT/m)$^2$ms. We emphasize that these parameters can be tailored to any specific application and may therefore result in more or less conservative restrictions on the waveform design.

*Experimental validation of Maxwell-compensated waveforms*

To verify that Maxwell-compensated waveforms (MCW) yield negligible error due to concomitant gradients we tested the waveforms by performing diffusion MRI in oil and in a healthy brain in vivo. To include a wide range of b-tensor shapes, we optimize waveforms that yield linear, planar and spherical b-tensors (LTE, PTE and STE, respectively) at b-values up to $b = 2.0$ ms/μm$^2$ while minimizing the encoding time (see Westin et al.[17] for definitions of b-tensor shapes).

All experiments were performed on a 3 T MAGNETOM Prisma (Siemens Healthcare, Erlangen, Germany) with maximal gradient amplitude of 80 mT/m and maximal slew rate of 200 T/m/s. We used a prototype pulse sequence developed in-house, based on the diffusion-weighted spin-echo sequence. The imaging parameters were TR = 4.5 s, FOV = 220×220×175 mm$^3$, slices = 35, resolution = 2×2×5 mm$^3$, iPAT = 2 (GRAPPA), echo spacing = 0.65 ms, partial-Fourier = 7/8, and b-values of 0.1, 0.7, 1.4 and 2.0 ms/μm$^2$. This setup yielded a separation between the first and second encoding periods of approximately 8 ms during which the refocusing pulse is executed. The EPI-readout started approximately 8 ms before the echo time such that $\delta_2 = \delta_1 - 8$ ms, where $\delta_1$ and $\delta_2$ are the durations of the first and second part of the waveform. This timing asymmetry was used for all waveform optimization. TE was set to 102 ms, as determined by the waveform that required the longest encoding time to reach the maximal b-value. We found that all waveforms could be robustly executed on the scanner, without violating duty cycle or peripheral nerve stimulation limits, by limiting the maximal gradient amplitude to 75 mT/m and the maximal slew rate to 60 T/m/s in the optimization and execution.

The reference waveform was a pair of symmetric trapezoidal gradient waveforms on either side of the refocusing pulse, i.e. the StejskalandTanner[34] design, also known as single diffusion encoding[35] that yields linear tensor encoding (SDE-LTE). For comparison, we also included non-compensated numerically optimized waveforms[5] that yield linear, planar and spherical encoding (NOW-LTE, NOW-PTE and NOW-STE). Finally, we included double diffusion encoding in an orthogonal configuration that yields planar tensor encoding (DDE-PTE). The DDE-PTE waveform required the longest encoding time, and therefore defined the TE that was used for all experiments. Note that all waveforms were executed in the timing for which they were optimized to retain the concomitant gradient characteristics that they exhibit at a minimal TE (Figure 3).

We predicted that none of the experiments would cause peripheral nerve stimulation above the permitted level. The prediction was performed in an in-house implementation of the SAFE model[36] and considered only the diffusion encoding gradient waveform and the EPI train.

A homogeneous oil phantom (cylinder with diameter 0.12 m, and length 0.20 m) was used to detect signal loss caused by concomitant gradients. Oil was chosen because it exhibits slow Gaussian diffusion ($< 0.01$ $\mu m^2$/ms). Thus, signal attenuation due to diffusion encoding is small (order of 1 %) so that substantial signal attenuation can be attributed to concomitant gradient effects[2]. We investigate the signal as a function of encoding strength for different waveform rotations and estimate the relative signal bias (Eq. 13) by comparing it to the reference sequence. Since the effect is expected to be rotation variant we use 16 rotations of the waveform for each b-value. The same rotations are used for each b-value. We use axial slice orientation ($\mathbf{n}_P = [0\ 1\ 0], \mathbf{n}_S = [0\ 0\ 1]$) and place the phantom so that the cylinder is approximately parallel to the main magnetic field. The FOV is placed so that the center of the outermost slices were at $z$-coordinates –25 and 125 mm. The FOV is shifted in this manner to probe signal errors at relatively large distances from the isocenter, because the magnitude of the concomitant fields increases with distance from the iso center.

Brain imaging was also performed in a healthy volunteer (age 38 y, male). The study was approved by the Regional Ethical Review Board (Lund, Sweden), and informed consent was obtained prior to participation. Each waveform was executed in 8, 12, 18 and 24 rotations for the four b-values, respectively. The diffusion encoding directions were generated by electrostatic repulsion on the half-sphere[37,38]. Each encoding direction corresponds to the symmetry axis of the b-tensors. Since spherical tensor encoding does not have a well-defined symmetry axis, we assign $g_1$ (Eq. 1) to be the symmetry axis. An axial slice orientation was used, and the center of the FOV was placed in the plane of the isocenter ($z = 0$), which coincided with the corpus callosum as seen in a mid-sagittal view. The center of the outermost slices covered $z$-coordinates –70 and 80 mm. The total scan time for all seven waveform designs was 36 min. Image volumes were corrected for motion and eddy currents in ElastiX[39], using extrapolated references[40].

To establish the size of the errors that may occur if concomitant gradients are ignored in the brain parenchyma, we investigate the impact on the diffusion-weighted signal at multiple locations. We manually defined regions of interest in the cerebellum, posterior (close to the isocenter) and anterior corpus callosum, and in the superior white matter close to the cortex. To take the diffusion anisotropy into account when comparing signals, we compare only identical b-tensors. Each comparison is between a non-compensated waveform and its Maxwell-compensated counterpart.

We also investigate the impact of concomitant gradients on parameters estimated by q-space trajectory imaging (QTI)[17,41]. QTI was fitted to data acquired with SDE-LTE combined with either compensated or non-compensated variants of planar tensor encoding (MCW-PTE or DDE-PTE, respectively). We estimated parameter maps of mean diffusivity (MD), fractional anisotropy (FA)[42], mean kurtosis (MK)[43,44], as well as the microscopic fractional anisotropy (µFA) and mean kurtosis decomposed into the isotropic ($MK_I$) and anisotropic ($MK_A$) components[17,20,23]. The parameter bias caused by concomitant gradients in the non-compensated case was calculated as the absolute difference in parameter values. To establish if the observed differences can be attributed to concomitant gradients, we predicted the bias in both signal and QTI parameters by assuming that the signal from MCW-PTE was free from errors and multiplying it by the attenuation factor estimated by Eq. 10 given the DDE-PTE waveform.

We have modified the numerical waveform optimization[5] to include Maxwell-compensation, and shared it at https://github.com/jsjol/NOW. We have also shared the SAFE peripheral nerve stimulation model at https://github.com/filip-szczepankiewicz/safe_pns_prediction, as well as tools for simulating and analyzing concomitant gradients and their effects as part of the multidimensional diffusion toolbox[45] at https://github.com/markus-nilsson/md-dmri. These tools were implemented in Matlab (The MathWorks, Natick, MA, USA).

*Simulated validation of Maxwell-compensated waveforms*

To verify that Maxwell-compensated waveforms yield negligible errors due to concomitant gradients for a comprehensive set of rotations, we simulate the expected signal bias according to Eq. 10 and Eq. 13. The simulations use the same waveforms and imaging settings as in the practical experiments. The worst signal bias within a sphere with a diameter of 0.2 m is calculated for three rotation modes. First, we investigate the effect of rotating the waveform assuming stationary axial slices. Second, the waveform is stationary but the FOV is rotated. Finally, both the waveform and FOV are rotated simultaneously but with independent rotation matrices. For each setup we use $10^4$ random rotations to yield a distribution of signal biases. We use boxplots to visualize the distribution of signal biases for each waveform and rotation mode.

*Impact of constraining the Maxwell index on encoding efficiency*

Constraining the Maxwell index in the waveform optimization is associated with a reduced efficiency of the diffusion encoding compared to optimization where the effects of concomitant gradients are ignored. We investigated how the maximal b-value depends on the Maxwell index by optimizing waveforms where *m* was constrained to values in the interval 1 to $10^5$ (mT/m)$^2$ms as well as optimization without constraining the Maxwell index. Thus, we span the range between extremely conservative to entirely unconstrained Maxwell indexes. The optimization settings were the same as for waveforms used in the practical experiments. The encoding efficiency is quantified in terms of the b-value that is achievable at a given constraint on the Maxwell-index relative to the maximal b-value, $b(m)/\max(b)$.

Since the Maxwell index reflects a balance of the time-integral of the gradients squared on both sides of the refocusing pulse, we also investigated how the efficiency depends on the encoding time asymmetry. The timing asymmetry was defined as $\delta_1/(\delta_1 + \delta_2)$, where $\delta_1$ and $\delta_2$ are the durations of the diffusion encoding before and after the refocusing pulse. The impact is also gauged in terms of the additional encoding time that is necessary to achieve the same b-value as for the non-compensated waveforms. We report the cost in terms of the additional encoding time required in absolute terms ($c_{\text{abs}} = t_{\text{MCW}} - t_{\text{NOW}}$) and relative terms ($c_{\text{rel}} = c_{\text{abs}}/t_{\text{NOW}}$) where $t$ is the total duration of Maxwell-compensated ($t_{\text{MCW}}$) and non-compensated ($t_{\text{NOW}}$) waveforms.

**Results**

Measurements of the diffusion-weighted signal in oil using the Maxwell-compensated waveforms appeared to be unaffected by concomitant gradients and had the same signal behavior as the reference sequence (Figure 3). At the highest b-value, the average bias across rotations (one standard deviation) was -0.1 (0.6), -0.2 (0.4) and -0.2 (0.5) in percent for linear, planar and spherical tensor encoding, respectively. By contrast, non-compensated waveforms yielded gross signal bias; the worst case was observed for DDE-PTE where some rotations exhibited a complete signal dropout (RB ≈ -100 %) at a distance of approximately 75 mm from the isocenter. Figure 3 also shows that the effect increases with increasing b-value, and that it introduces rotational variance even if the diffusion itself is perfectly isotropic. Overall, the phantom experiments consistently verified that the Maxwell index is an appropriate target for optimization to alleviate the effects of concomitant gradients, and that ignoring such effects may have a marked effect on the diffusion-weighted signal.

The effects of concomitant gradients were also investigated in the brain in vivo. Figure 4 shows that non-compensated waveforms can suffer effects that are large enough to affect the average signal, whereas the Maxwell-compensated waveforms exhibit no evident attenuation due to concomitant gradients. As in the oil, the effect depends on the rotation of the waveform and is larger for higher b-values. Therefore, the impact on parameters that correspond to the initial slope of *S*(*b*), such as the mean diffusivity, is likely smaller compared to parameters based on signal characteristics at high b-values, such as microscopic anisotropy and kurtosis parameters.

The diffusion-weighted signal and QTI parameter maps for Maxwell-compensated and non-compensated waveforms are shown in Figure 5. The diffusion-weighted signal for DDE-PTE at $b$ = 2 ms/µm² is visibly affected by signal dropout, most prominent along the feet-head direction, whereas MCW-PTE exhibited no such behavior. We emphasize that the top row of the figure only shows the signal map for a single rotation of the waveform, and that the artifacts produced by DDE-PTE take on different appearances depending on the rotation, as seen in the oil phantom where *S*(*b*) varies wildly depending on the waveform rotation (Figure 3). As expected, the differences in MD and FA were relatively small, but gross parameter differences were observed in the μFA, $MK_A$ (overestimated by 0.3 and 0.8, respectively) and $MK_I$ (underestimated by 0.8). The measured and predicted differences show a remarkable resemblance, indicating that the difference between signal and QTI parameters based on DDE-PTE and MCW-PTE indeed differ primarily due to the prominent concomitant gradients created by DDE-PTE.

Simulations presented in Figure 6 show that the Maxwell-compensated waveforms can retain a negligible signal bias (magnitude is always below 1 %) due to concomitant gradients regardless of rotations of the waveform and/or FOV. By contrast, all non-compensated waveforms show a large variation of the bias depending on rotations.

The cost of imposing a restriction on the Maxwell index in the numerical optimization of the waveforms is that the encoding efficiency, in terms of the b-value, is reduced. Using the current premise, the b-value was reduced by 7–23 %, depending on b-tensor shape (Figure 7). This translates to an extension of the required encoding time to reach $b$ = 2 ms/μm² of approximately $c_{abs}$ = 1–6 ms or $c_{rel}$ = 2–8 %. Furthermore, the efficiency penalty depends on the sequence timing asymmetry, where equal timing on both sides of the refocusing pulse yields the highest efficiency.

**Discussion**

It is critical to consider signal errors and artifacts caused by concomitant gradients in diffusion encoding with asymmetric waveforms. We have proposed the Maxwell index as a target for waveform optimization and demonstrated that waveforms with low Maxwell indices (100–1000 (mT/m)²ms) exhibit negligible effects of concomitant gradients. This result generalizes to arbitrary rotations of the gradient waveform (encoding directions) and FOVs with arbitrary size and orientation. As such, the optimization is performed only once— under appropriate conditions—and will produce a waveform that will have negligible bias due to concomitant gradients, independent of the exact setup of future experiments. Additionally, we have presented a compact matrix formulation for the approximation of concomitant gradients for an arbitrary number of refocusing pulses and expanded the error prediction for slice selective spin-echo experiments to include the effect of $T_2^*$-relaxation.

Investigations of concomitant gradient effects revealed that the impact of concomitant gradients is large for several asymmetric gradient waveform designs proposed in literature. Most prominently, DDE in a spin-echo can be used to showcase the gross errors that can appear due to concomitant gradients. Figures 3 and 5 show massive signal loss (RB ≈ -100 %) both in a phantom and in the human brain. Although we cannot rule out that the signal in the brain depends also on effects that are not captured by the b-tensor, we expect that diffusion time dependency, exchange and flow are negligible compared to the gross signal loss that is predicted by theory and seen in the experiments when using non-compensated waveforms. Thus, it is reasonable to assume that the differences in diffusion-weighted signal and parameter maps in Figure 5 can be attributed mainly to a marked signal bias caused by concomitant gradients when using DDE-PTE. Notably, MCW-PTE is a viable alternative to DDE-PTE, since it is both robust to concomitant gradient effects *and* more efficient (shorter minimal TE for any given b-value). We stress that this issue with DDE is not present for imaging sequences where multiple refocusing pulses are used to make the diffusion encoding symmetric, such as in the double-spin-echo[46].

Maxwell-compensated waveforms are compatible with existing approaches for alleviating the effects of concomitant gradients. For example, it is possible to subtract the concomitant waveform for a given position (e.g., center of the slice) from the desired waveform such that the actual waveform is closer to the desired waveform, as suggested by Meier et al.[4]. The drawback of this approach is that the correction holds only for one point in space, whereas remaining parts of the slice still suffer some error that may be non-negligible. Another drawback is that such single-slice corrections may be incompatible with techniques that excite multiple slices simultaneously[47], although adaptions have been proposed[48]. Another approach is post-hoc correction of the signal data by dividing with the predicted attenuation factor in Eq. 10. This approach is limited by its requirement for very detailed knowledge of the sequence design as well as the $T_2^*$ of the tissue. Furthermore, if the error is so large that most of the signal is lost (Figure 3 and Figure 5) the impact on SNR will be significant and unavoidable. Thus, we argue that Maxwell-compensated waveforms may be preferable to post-hoc correction, especially for diffusion encoding at high gradient amplitudes and/or long encoding times[49,50], imaging of organs in large FOVs, super-resolution methods that employ thick slices and/or techniques based on rotated/oblique imaging stacks[51-54].

The drawback of Maxwell-compensation is a penalty to the encoding efficiency which manifests as a slightly longer encoding time. However, the penalty is rather small (1–6 ms), at least for the conditions used in this paper. Notably, the penalty grows with increasing timing asymmetry (Figure 7), although we note that this is true for symmetric designs too. Asymmetric designs that ignore concomitant gradients can be virtually independent of timing asymmetry, but the signal bias is expected to grow with increasing asymmetry.

The premise of this study is limited to that of brain imaging at a 3 T clinical MRI system with b-values up to 2 ms/µm$^2$. As such, we do not cover other relevant scenarios in detail. For example, we assume a $T_2^*$ of 40 ms to simulate tissue, which may not always be a representative. Thus, specific results of this study may not be immediately translatable to other systems, sequence designs, and tissues. However, we have shared the theoretical framework, waveform optimization and analysis software in open source so that it can be used to develop and test waveform and experiment designs in other scenarios. Indeed, it can be used to decide if Maxwell-compensation is at all necessary, or if the premise allows for the use of non-compensated waveforms due to, for example, very small FOVs. Furthermore, the theory in this work is limited to estimating the linear spatial terms of the concomitant gradients—assuming that higher order terms are negligible[1]—and is valid only for MRI systems that have symmetric gradient coils[4].

**Conclusions**

In conclusion, we have proposed and demonstrated that the Maxwell index is a valid optimization target in the design of asymmetric gradient waveforms to yield negligible concomitant gradient effects. The design is applicable to arbitrary b-tensor shapes, sequence timing, and for any number of refocusing pulses larger than zero. The waveforms are Maxwell-compensated throughout the imaging volume independent of waveform rotation and FOV orientation. As such, this approach is especially relevant for diffusion imaging at high gradient strength, long encoding times, low main magnetic field, large/oblique/off-isocenter FOVs, thick slices, and simultaneous multi-slice acquisition.

## Acknowledgements

We thank Siemens Healthcare for providing source code and the pulse sequence programming environment. We thank Dr. Thomas Witzel for developing the Matlab implementation of the SAFE-model. FS and MN are inventors on patents related to the study. MN declares research support from and ownership interest in Random Walk Imaging, which holds patents related to this study. This study was supported by the Swedish Research Council (grants no. 2016-03443, and 2016-04482), Swedish Foundation for Strategic Research (grant no. AM13-0090), Random Walk Imaging AB (grant no. MN15), and NIH (grant no. R01MH074794 and P41EB015902).

**Figures**

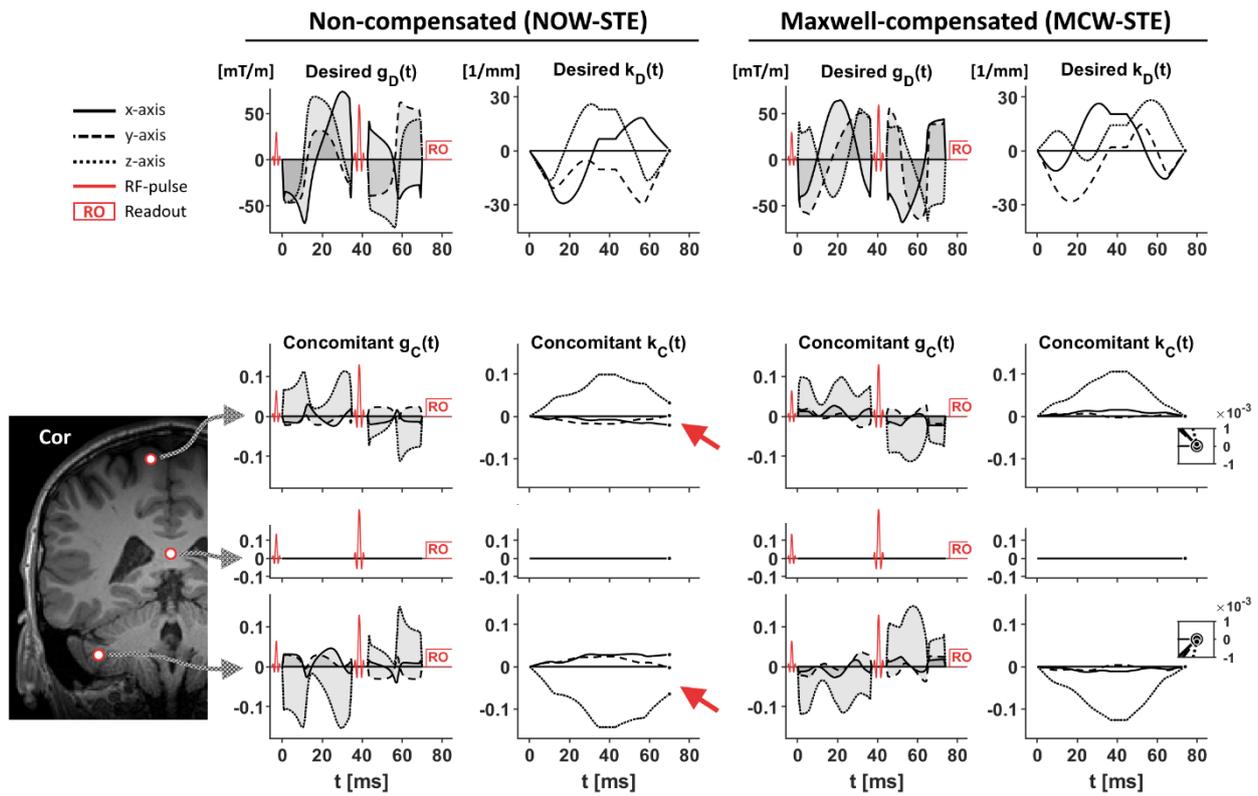

Figure 1 – Desired gradient waveforms and their concomitant gradients at the three positions in the brain. Top row shows the desired gradient waveforms—non-compensated (left) and Maxwell-compensated (right)—and their moment vectors as a function of time. The intension is that the desired moment $\mathbf{k}_D = 0$ at the end of the diffusion encoding, which is a trivial condition to fulfil in the absence of concomitant gradients. However, concomitant gradients will always appear within the object to varying degree. The concomitant gradients are not strong enough to cause a relevant error in the b-tensor, but may cause a residual gradient moment that persists after the end of the diffusion encoding (red arrows). The size and direction of the residual moment vector depends on the position in space, increasing in magnitude further away from the isocenter. By contrast, the Maxwell-compensated waveforms are designed to have concomitant gradients that exhibit a negligible residual moment. The inset plots in the rightmost column show that the residual moment after the diffusion encoding is vanishingly small. The three positions (**r**) evaluated in this example are—from top to bottom—[10 0 60] mm, [0 0 0] mm (isocenter) and [50 0 –80] mm.

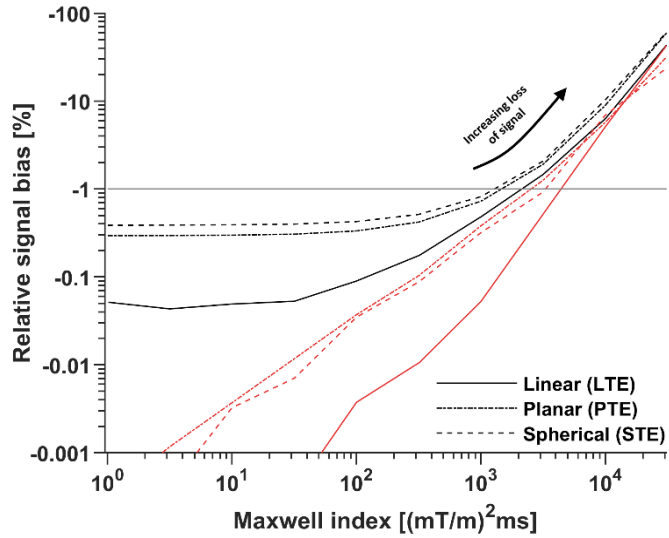

Figure 2 – The relative signal bias depends on the Maxwell index constraint as well as the target b-tensor shape. The plot shows the simulated bias in the worst-case scenario when using interpolated (black) and original (red) waveforms optimized at a temporal resolution of $N = 100$ samples. It is clear that the interpolation worsens the Maxwell-compensation but does not elevate the error above the tolerated level of 1 % (gray line). Furthermore, the signal bias becomes virtually independent of the Maxwell index when $m$ is below approximately 100 $(mT/m)^2 ms$. This indicates diminishing returns for overly conservative optimization of $m$ when the waveform must be interpolated in a later step.

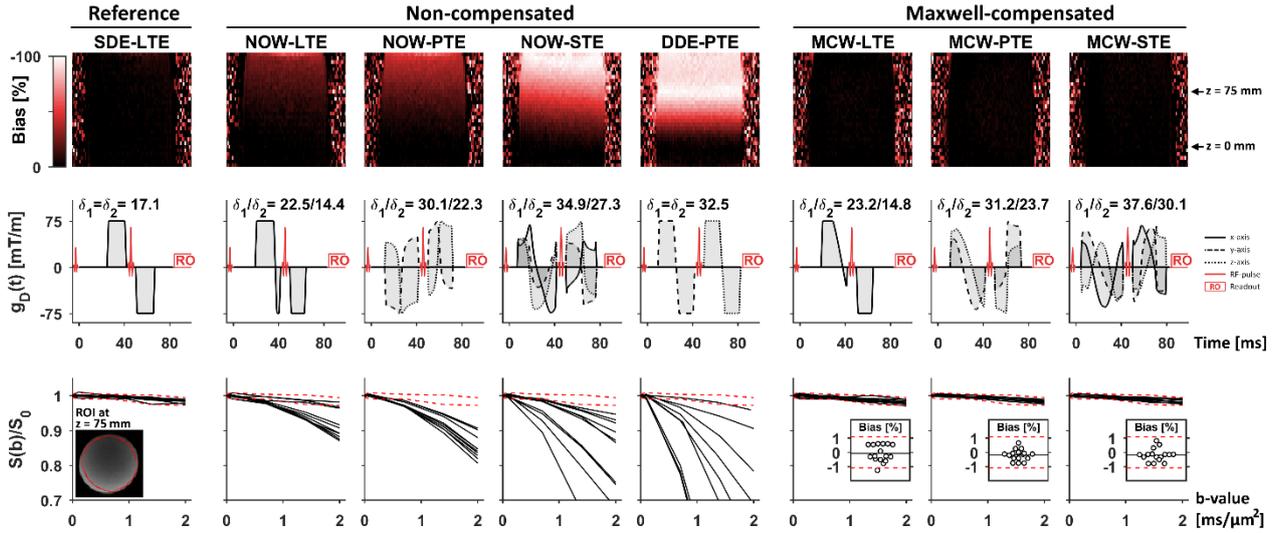

Figure 3 – Diffusion-weighted signal in an oil phantom shows that non-compensated asymmetric gradient waveform designs exhibit gross signal error due to concomitant gradients, whereas the Maxwell-compensated waveforms exhibit negligible errors. The top row shows the relative signal bias (Eq. 13) in coronal slices at the maximal b-value for the worst waveform rotation that was observed for each waveform type. The middle row shows the waveforms and timing used in the experiments. The bottom row shows the average signal in an ROI placed at $z = 75$ mm from the isocenter as a function of b-value for 16 rotations; broken red lines show the interval spanned by the average reference signal ± two standard deviations. The signal bias is worst for DDE-PTE, reaching approximately –100 % (AF ≈ 0) in a large part of the volume. Bias also appears for remaining non-compensated waveforms, albeit to a lesser degree. By contrast, the Maxwell-compensated waveforms showed no sign of concomitant gradient effects. The inset plots in the bottom row show the estimated signal bias in the ROI at $b = 2$ ms/µm$^2$ where horizontal black lines show the average bias across all rotations. All Maxwell-compensated waveforms had negligible signal bias and the distribution across rotations exhibited a variability that was similar to the reference sequence.

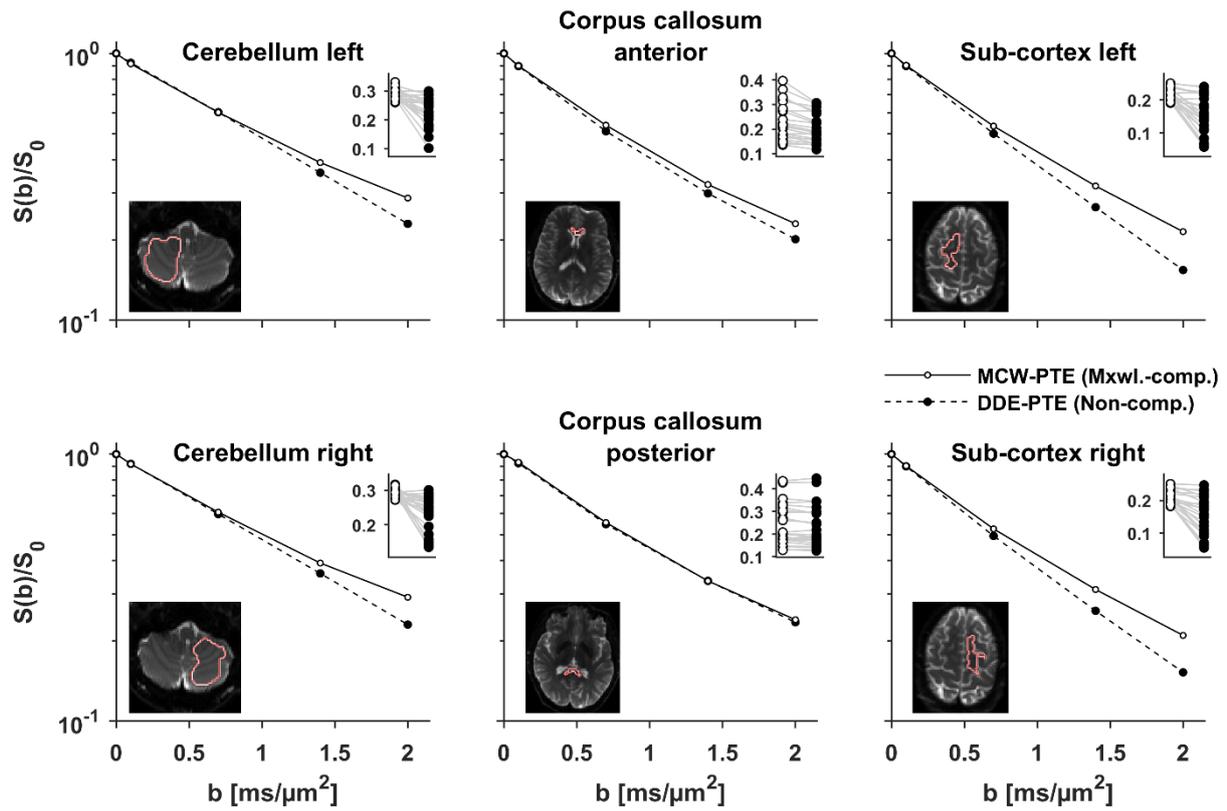

Figure 4 – Plots show the signal vs b-value for planar tensor encoding produced by non-compensated (DDE-PTE, black markers) and Maxwell-compensated waveforms (MCW-PTE, white markers). In the main plots, the signal is averaged over directions, whereas the inset plots show individual directions at $b = 2$ ms/µm$^2$ where gray lines link signal pairs encoded with identical b-tensors. At low b-values the two waveforms are similar, however at larger b-values, DDE-PTE suffers gross signal attenuation due to concomitant gradients. The effect is largest in the inferior and superior parts of the brain, i.e. the bias is largest along the z-axis. As expected from theory, the effect is negligible in the posterior corpus callosum because of its proximity to the isocenter.

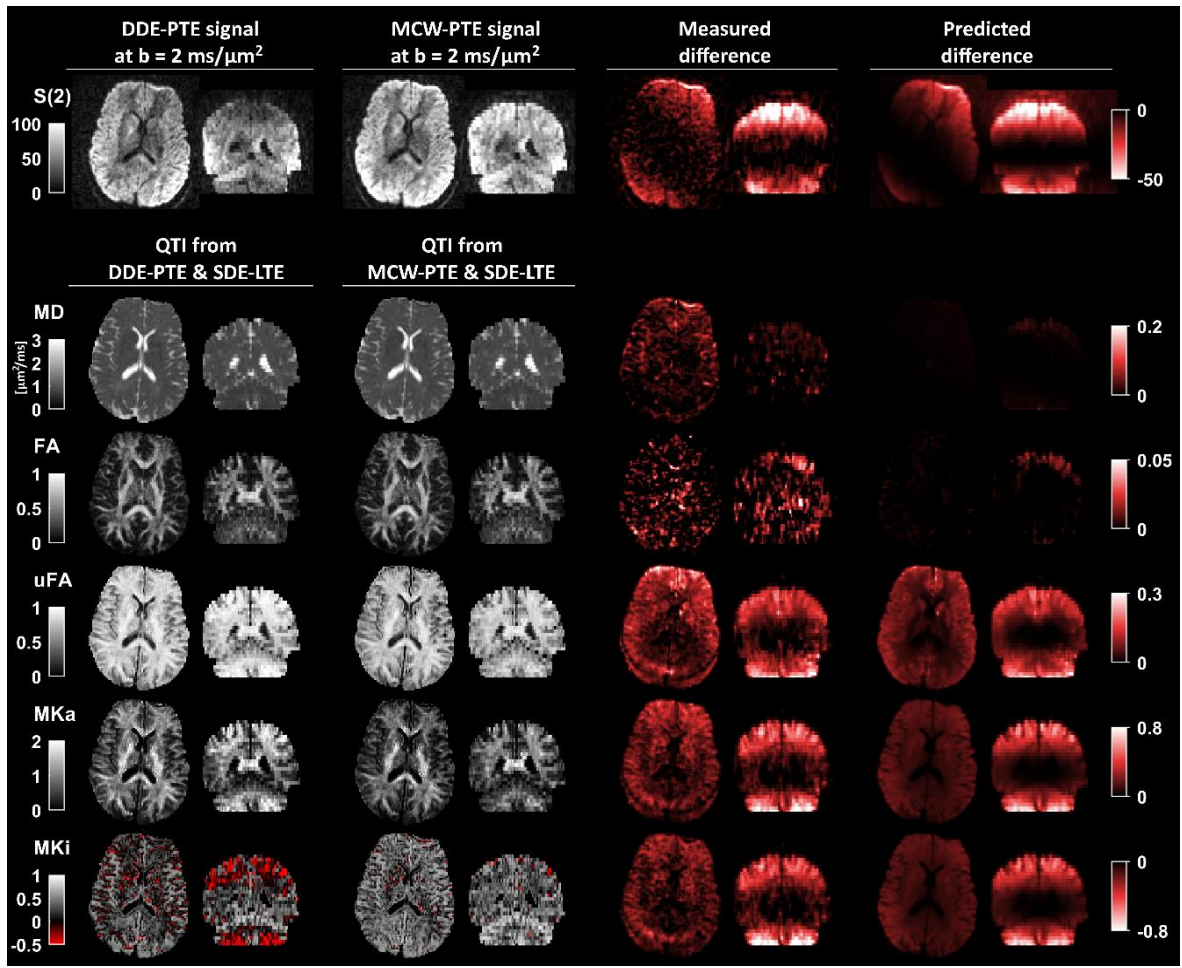

Figure 5 – Diffusion-weighted signal maps and QTI parameter maps calculated from a combination of linear and planar tensor encoding. The linear encoding used the symmetric Stejskal-Tanner design (SDE-LTE), whereas the planar encoding is achieved by either non-compensated (DDE-PTE) or Maxwell-compensated (MCW-PTE) waveforms. There are striking differences in the signal using planar encoding at $b = 2$ ms/µm$^2$ (top row), as well as in several parameter maps. The top row shows that the signal bias can be pronounced both within and between slices. The axial and coronal slices show examples where DDE-PTE exhibited a marked loss of signal in the anterior/posterior and inferior/superior regions, while images acquired with MCW-PTE had no visible artifacts or signal loss due to concomitant gradients. Since it is only the planar encoding that is affected by concomitant gradients, the divergence between the linear and planar encoding increases artificially. This explains why DDE-PTE causes a severe overestimation of µFA and MK$_A$, and underestimation of the MK$_I$[20,28]. The bias is severe enough to causes the MK$_I$ to be negative in large portions of the brain. The measured and predicted differences were similar in both magnitude and overall geometry, which indicates that they are caused primarily by concomitant gradients in the DDE-PTE.

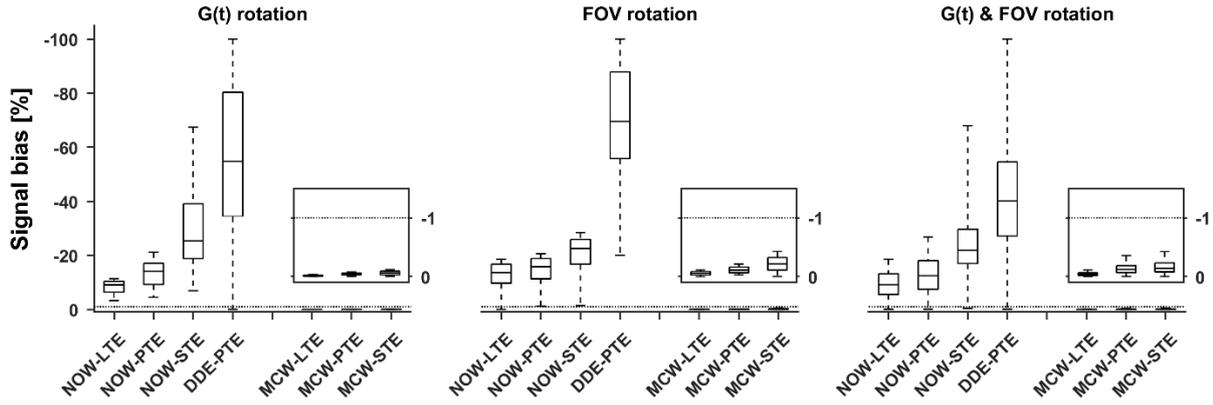

Figure 6 – Simulation of signal bias for Maxwell-compensated and non-compensated waveforms that yield linear (LTE), planar (PTE) and spherical (STE) b-tensor encoding. The boxplots show the distribution of signal bias under $10^4$ random rotations of the waveform, FOV, or both; whiskers show extreme values, the box lines show 25, 50 and 75 percentiles. The simulated waveforms are identical to those used in the practical experiments (Figure 3) and scaled to yield $b = 2$ ms/µm$^2$. The non-compensated waveforms suffer from significant rotation dependent bias. Notably, DDE-PTE exhibits the worst overall bias. Throughout all examples the Maxwell-compensated waveforms exhibit negligible bias, regardless of b-tensor shape and rotation scheme; the simulation predicts that both the waveform and FOV can be arbitrarily rotated without discernible signal bias. The inset plots show a magnification of the bias for the Maxwell-compensated waveforms (y-axis covers the interval 0 to -1.5 %). The dotted line shows the level where 1 % of the signal is lost due to concomitant gradient effects.

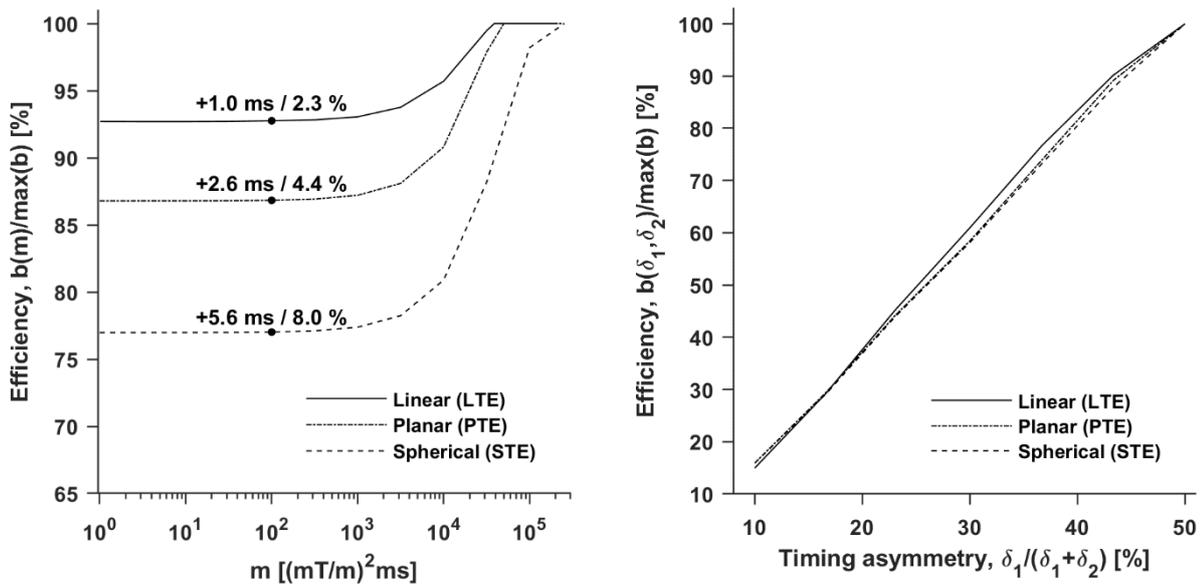

Figure 7 – Introducing a limit on the Maxwell index reduces the efficiency of the diffusion encoding and the penalty depends on the timing asymmetry. The left plot shows that the impact on efficiency is independent of the Maxwell index for *m* below approximately 1000 (mT/m)²ms. The absolute and relative impact on the encoding duration ($c_{abs}$ / $c_{rel}$) is stated for the experiment setup used in the practical experiments ($\delta_2 = \delta_1 -$ 8 ms) at *m* = 100 (mT/m)²ms. For completeness, the same penalty is 3.6 ms, or 7.9 %, for STE when using max-norm optimization (data not shown). The right plot shows that the efficiency depends heavily on the encoding time asymmetry of the sequence. As expected, equal gradient duration on either side of the refocusing pulse yield the highest efficiency.

**References**

1. Bernstein MA, Zhou XJ, Polzin JA, et al. Concomitant gradient terms in phase contrast MR: analysis and correction. *Magn Reson Med.* 1998;39(2):300-308.
2. Baron CA, Lebel RM, Wilman AH, Beaulieu C. The effect of concomitant gradient fields on diffusion tensor imaging. *Magn Reson Med.* 2012;68(4):1190-1201.
3. Szczepankiewicz F, Nilsson M. Maxwell-compensated waveform design for asymmetric diffusion encoding. Paper presented at: Proc. Intl. Soc. Mag. Reson. Med. 26.2018; Paris, France.
4. Meier C, Zwanger M, Feiweier T, Porter D. Concomitant field terms for asymmetric gradient coils: consequences for diffusion, flow, and echo-planar imaging. *Magn Reson Med.* 2008;60(1):128-134.
5. Sjölund J, Szczepankiewicz F, Nilsson M, Topgaard D, Westin CF, Knutsson H. Constrained optimization of gradient waveforms for generalized diffusion encoding. *J Magn Reson.* 2015;261:157-168.
6. Hutter J, Tournier JD, Price AN, et al. Time-efficient and flexible design of optimized multishell HARDI diffusion. *Magn Reson Med.* 2018;79(3):1276-1292.
7. Aliotta E, Moulin K, Ennis DB. Eddy current-nulled convex optimized diffusion encoding (EN-CODE) for distortion-free diffusion tensor imaging with short echo times. *Magn Reson Med.* 2018;79(2):663-672.
8. Hargreaves BA, Nishimura DG, Conolly SM. Time-optimal multidimensional gradient waveform design for rapid imaging. *Magn Reson Med.* 2004;51(1):81-92.
9. Laun FB, Kuder TA. Diffusion pore imaging with generalized temporal gradient profiles. *Magn Reson Imaging.* 2013;31(7):1236-1244.
10. Demberg K, Laun FB, Bertleff M, Bachert P, Kuder TA. Experimental determination of pore shapes using phase retrieval from q-space NMR diffraction. *Phys Rev E.* 2018;97(5-1):052412.
11. Norris DG, Hutchison JMS. Concomitant magnetic field gradients and their effects on imaging at low magnetic field strengths. *Magn Reson Imaging.* 1990;8(1):33-37.
12. Zhou XJ, Du YP, Bernstein MA, Reynolds HG, Maier JK, Polzin JA. Concomitant magnetic-field-induced artifacts in axial echo planar imaging. *Magn Reson Med.* 1998;39(4):596-605.
13. Du YP, Joe Zhou X, Bernstein MA. Correction of concomitant magnetic field-induced image artifacts in nonaxial echo-planar imaging. *Magn Reson Med.* 2002;48(3):509-515.
14. Irfanoglu MO, Walker L, Sarlls J, Marenco S, Pierpaoli C. Effects of image distortions originating from susceptibility variations and concomitant fields on diffusion MRI tractography results. *Neuroimage.* 2012;61(1):275-288.
15. Testud F, Gallichan D, Layton KJ, et al. Single-shot imaging with higher-dimensional encoding using magnetic field monitoring and concomitant field correction. *Magn Reson Med.* 2015;73(3):1340-1357.
16. Zhou XJ, Tan SG, Bernstein MA. Artifacts induced by concomitant magnetic field in fast spin-echo imaging. *Magn Reson Med.* 1998;40(4):582-591.
17. Westin CF, Knutsson H, Pasternak O, et al. Q-space trajectory imaging for multidimensional diffusion MRI of the human brain. *Neuroimage.* 2016;135:345-362.
18. Topgaard D. Multidimensional diffusion MRI. *J Magn Reson.* 2017;275:98-113.
19. Mitra P. Multiple wave-vector extensions of the NMR pulsed-field-gradient spin-echo diffusion measurement. *Physical Review B.* 1995;51(21):15074-15078.
20. Lasič S, Szczepankiewicz F, Eriksson S, Nilsson M, Topgaard D. Microanisotropy imaging: quantification of microscopic diffusion anisotropy and orientational order parameter by diffusion MRI with magic-angle spinning of the q-vector. *Frontiers in Physics.* 2014;2:11.
21. Jespersen SN, Lundell H, Sønderby CK, Dyrby TB. Orientationally invariant metrics of apparent compartment eccentricity from double pulsed field gradient diffusion experiments. *NMR Biomed.* 2013;26(12):1647-1662.
22. Lawrenz M, Finsterbusch J. Detection of microscopic diffusion anisotropy on a whole-body MR system with double wave vector imaging. *Magn Reson Med.* 2011;66(5):1405-1415.
23. Szczepankiewicz F, Lasič S, van Westen D, et al. Quantification of microscopic diffusion anisotropy disentangles effects of orientation dispersion from microstructure: Applications in healthy volunteers and in brain tumors. *Neuroimage.* 2015;104:241-252.

24. Shemesh N, Ozarslan E, Adiri T, Basser PJ, Cohen Y. Noninvasive bipolar double-pulsed-field-gradient NMR reveals signatures for pore size and shape in polydisperse, randomly oriented, inhomogeneous porous media. *J Chem Phys.* 2010;133(4):044705.
25. Özarslan E, Basser PJ. Microscopic anisotropy revealed by NMR double pulsed field gradient experiments with arbitrary timing parameters. *J Chem Phys.* 2008;128(15):154511.
26. Avram AV, Ozarslan E, Sarlls JE, Basser PJ. In vivo detection of microscopic anisotropy using quadruple pulsed-field gradient (qPFG) diffusion MRI on a clinical scanner. *Neuroimage.* 2013;64:229-239.
27. Nilsson M, Englund E, Szczepankiewicz F, van Westen D, Sundgren PC. Imaging brain tumour microstructure. *Neuroimage.* 2018.
28. Szczepankiewicz F, van Westen D, Englund E, et al. The link between diffusion MRI and tumor heterogeneity: Mapping cell eccentricity and density by diffusional variance decomposition (DIVIDE). *Neuroimage.* 2016;142:522-532.
29. Lampinen B, Szczepankiewicz F, Martensson J, van Westen D, Sundgren PC, Nilsson M. Neurite density imaging versus imaging of microscopic anisotropy in diffusion MRI: A model comparison using spherical tensor encoding. *Neuroimage.* 2017;147:517-531.
30. Mori S, van Zijl P. Diffusion Weighting by the Trace of the Diffusion Tensor within a Single Scan. *Magn Reson Med.* 1995;33(1):41-52.
31. Wong EC, Cox RW, Song AW. Optimized isotropic diffusion weighting. *Magn Reson Med.* 1995;34(2):139-143.
32. Eriksson S, Lasič S, Topgaard D. Isotropic diffusion weighting in PGSE NMR by magic-angle spinning of the q-vector. *J Magn Reson.* 2013;226:13-18.
33. Cory DG, Garroway AN, Miller JB. Applications of Spin Transport as a Probe of Local Geometry. *Abstr Pap Am Chem S.* 1990;199:105.
34. Stejskal EO, Tanner JE. Spin Diffusion Measurement: Spin echoes in the Presence of a Time-Dependent Field Gradient. *the journal of chemical physics.* 1965;42(1):288-292.
35. Shemesh N, Jespersen SN, Alexander DC, et al. Conventions and nomenclature for double diffusion encoding NMR and MRI. *Magn Reson Med.* 2016;75(1):82-87.
36. Hebrank FX, Gebhardt M. SAFE-Model - A New Method for Predicting Peripheral Nerve Stimulations in MRI. Paper presented at: Proc. Intl. Soc. Mag. Res. Med.2000.
37. Jones DK, Horsfield MA, Simmons A. Optimal strategies for measuring diffusion in anisotropic systems by magnetic resonance imaging. *Magn Reson Med.* 1999;42(3):515-525.
38. Leemans A. ExploreDTI: a graphical toolbox for processing, analyzing, and visualizing diffusion MR data. Paper presented at: Proc. Intl. Soc. Mag. Reson. Med.2009.
39. Klein S, Staring M, Murphy K, Viergever MA, Pluim JP. elastix: a toolbox for intensity-based medical image registration. *IEEE Trans Med Imaging.* 2010;29(1):196-205.
40. Nilsson M, Szczepankiewicz F, van Westen D, Hansson O. Motion and eddy-current correction in high b-value diffusion MRI: Systematic registration errors and how to avoid them. Paper presented at: Proc. Intl. Soc. Mag. Reson. Med.2015.
41. Westin CF, Szczepankiewicz F, Pasternak O, et al. Measurement tensors in diffusion MRI: Generalizing the concept of diffusion encoding. *Med Image Comput Comput Assist Interv.* 2014;17 (Pt 5):217-225.
42. Basser PJ, Mattiello J, Le Bihan D. MR diffusion tensor spectroscopy and imaging. *Biophys J.* 1994;66(1):259-267.
43. Jensen JH, Helpern JA, Ramani A, Lu H, Kaczynski K. Diffusional kurtosis imaging: the quantification of non-gaussian water diffusion by means of magnetic resonance imaging. *Magn Reson Med.* 2005;53(6):1432-1440.
44. Yablonskiy DA, Bretthorst GL, Ackerman JJ. Statistical model for diffusion attenuated MR signal. *Magn Reson Med.* 2003;50(4):664-669.
45. Nilsson M, Szczepankiewicz F, Lampinen B, et al. An open-source framework for analysis of multidimensional diffusion MRI data implemented in MATLAB. Paper presented at: Proc. Intl. Soc. Mag. Reson. Med. 262018; Paris, France.
46. Callaghan PT, Komlosh ME. Locally anisotropic motion in a macroscopically isotropic system: displacement correlations measured using double pulsed gradient spin-echo NMR. *Magnetic Resonance in Chemistry.* 2002;40(13):S15-S19.

47. Larkman DJ, Hajnal JV, Herlihy AH, Coutts GA, Young IR, Ehnholm G. Use of multicoil arrays for separation of signal from multiple slices simultaneously excited. *J Magn Reson Imaging.* 2001;13(2):313-317.
48. Bhat H, Hoelscher U, Zeller M, Inventors. METHOD AND MAGNETIC RESONANCE APPARATUS FOR MAXWELL COMPENSATION IN SIMULTANEOUS MULTISLICE DATA ACQUISITIONS. 2017.
49. Jones DK, Alexander DC, Bowtell R, et al. Microstructural imaging of the human brain with a 'super-scanner': 10 key advantages of ultra-strong gradients for diffusion MRI. *Neuroimage.* 2018;182:8-38.
50. Does MD, Parsons EC, Gore JC. Oscillating gradient measurements of water diffusion in normal and globally ischemic rat brain. *Magn Reson Med.* 2003;49(2):206-215.
51. Scherrer B, Gholipour A, Warfield SK. Super-resolution reconstruction to increase the spatial resolution of diffusion weighted images from orthogonal anisotropic acquisitions. *Med Image Anal.* 2012;16(7):1465-1476.
52. Setsompop K, Fan Q, Stockmann J, et al. High-resolution in vivo diffusion imaging of the human brain with generalized slice dithered enhanced resolution: Simultaneous multislice (gSlider-SMS). *Magn Reson Med.* 2017.
53. Van Steenkiste G, Jeurissen B, Veraart J, et al. Super-resolution reconstruction of diffusion parameters from diffusion-weighted images with different slice orientations. *Magn Reson Med.* 2016;75(1):181-195.
54. Jeurissen B, Sijbers J, Szczepankiewicz F. Improved precision and accuracy in q-space trajectory imaging by model-based superresolution reconstruction. Paper presented at: Proc. Int. Soc. Magn. Reson. Med. 272019.